\documentclass[]{mn2e}
\usepackage{epsfig}
\usepackage{amssymb}
\begin{document}
\title[A regularized free form estimator for dark energy]
{A Regularized Free Form Estimator for Dark Energy}
{
\author[T D Saini]
{Tarun Deep Saini\\
Institute of
Astronomy, University of Cambridge, Madingley Road, Cambridge CB3
0HA.\\
} 
%
%
\maketitle
\begin{abstract}
We construct a simple, regularized estimator for the dark energy
equation of state by using the recently introduced linear response
approximation. We show that even a simple regularization substantially
improves the performance of the free form fitting approach.  The use
of linear response approximation allows an analytic construction of
maximum likelihood estimator, in a convenient and easy to use matrix
form. We show that in principle, such regularized free form fitting
can give us an unbiased estimate of the functional form of the
equation of state of dark energy.  We show the efficacy of this
approach on a simulated SNAP class data, but it is easy to generalize
this method to include other cosmological tests.  We provide a
possible explanation for the sweet spots seen in other reconstruction
methods.

\end{abstract}
\begin{keywords}
cosmology:theory -- methods: statistical --cosmological parameters.
\end{keywords}
\section{Introduction}

One of the most exciting discovery of the last decade is the
possibility that the expansion of our Universe is accelerating (Perlmutter
et al.~1999; Riess et al.~1998). The simplest explanation in terms of
a cosmological constant runs into to a fine tuning problem (Sahni \&
Starobinski 2000; Peebles \& Ratra, 2002; Padmanabhan, 2002).
Therefore, it has become popular to phenomenologically model the
component that drives the acceleration as an ideal fluid with an
equation of state given by $P= w\rho$, where the equation of state
parameter $w$ is allowed to vary with time. In this parameterization
the cosmological constant model corresponds to $w=-1$. In the recent
years there has been a considerable interest in devising methods to
extract information about the equation of state from the present and
the future cosmological data.

In principle, since this information is coded directly into the
distance (luminosity and angular) measures, it is possible to directly
obtain $w(z)$ from the supernovae distances (Starobinsky, 1998).  This
requires the knowledge of up to a third derivative of noisy estimate
of the distance measures with respect to the redshift. This makes such
direct estimation extremely noisy. Methods based on flexible fitting
functions (Saini et al. 2000; Nakamura \& Chiba 2001) for the
luminosity distance have been invoked to get around this problem. In
such schemes the number of parameters in the fitting function is kept
small, therefore, the allowed behaviour of the equation of state is
restricted by the adoption of specific forms for the fitting
functions. Other popular methods approximate the dark energy density
or the unknown function $w(z)$ as low order polynomials (Sahni et
al.~2002; Weller \& Albrecht,~2002).  Since the quantity of interest
is the equation of state, direct expansion of $w(z)$ are better able
to constrain the dark energy.  Saini et al. (2003) (hereafter SPB)
show that the distance measures are approximately linear functionals
of the equation of state in the possible range of parameters. Using
this they calculate the expectation value of the polynomial
approximations for any given $w(z)$. They conclude that schemes based
on polynomial expansion of the equation of state are useful since they
measure certain well defined, integrated properties of the underlying,
true equation of state.

Wang \& Lovelace (2001) show that by considering the dark energy
density in redshift bins the bias inherent in the finite
parameterization of the dark energy can be easily removed. Huterer \&
Starkman (2002) apply a similar method to binned equation of state
parameter, $w(z)$. A limitation of this approach is that due to the
large number of bins required to reconstruct the precise behaviour of
the dark energy, the estimated equation of state turns out to be very
noisy. Huterer \& Starkman (2002) find that although a direct
reconstruction looks hopelessly noisy, useful information about the
equation of state is still coded into the Fisher matrix. They show
that by diagonalizing it, a few principle components could still be
measured with good accuracy from the future experiments. They advocate
the use of eigen vectors of the Fisher matrix as the optimal basis to
express the unknown function $w(z)$.

Such free form reconstructions of the equation of state do not return
a smooth function. Our main aim in this paper is to show that simple
regularization of the free form estimation helps substantially in
bringing down the noise in the reconstruction. In SPB the linear
response approximation was used to formulate a similar free form
reconstruction in a matrix form.  We use this approximation in this
paper to investigate the effect of introducing smoothness as a
constraint. Although the accuracy of the linear response approximation
is limited, we make use of it in this paper since it enables the
regularization to be done analytically. Similar results hold for the
exact case. We also show that an iterative scheme could in principle
extend the applicability of the linear response approximation, while
still retaining all the advantages of its analytical simplicity.

The plan of this paper is as follows. In Section~2 we collate the
results on linear response approximation.  In Section~3 we formulate a
discrete version, more suited to real data and apply it to
construct the free form estimator for the equation of state.  We then
add the regularizing terms to modify the estimator to guarantee
smoothness.  In Section~4 we use a simple model for dark energy to
illustrate the performance of the regularized estimation and compare
it to the unregularized one.  Our conclusions are presented in
Section~5.
\begin{figure}
\vbox{\center{\centerline{
\mbox{\epsfig{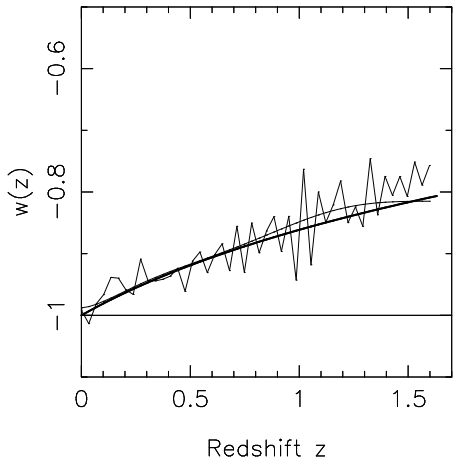}}
}}}
\caption{A free form reconstruction of $w(z)=-1 + 0.2 \ln(1+z)$ without 
added noise. The thick solid line shows the input model, the wiggly
line is the reconstruction without any regularization and the solid
line shows reconstruction with regularization. Small departures form
the true equation of state are due to numerical noise and inexactness
of the linear response approximation.  }
\label{fig:figure1}
\end{figure}
\section{Linearized luminosity distance}
\label{sec:funct}

The exact relation between the luminosity distance and the equation of
state $w(z)$ is non-linear, however, it was shown in SPB that by
considering a given equation of state as small departure from a
fiducial equation of state we can approximately linearize this
relation.  In this section we collate the necessary expressions.
In a spatially flat universe the luminosity distance is given by
\begin{equation}
D_L(z) = (1+z)(1+g)^{1/2} \,\int_1^{1+z} dx\, \frac{x^{-3/2}}
{\left [ g + Q[w,x] \right ]^{1/2}}\,\,,
\label{eq:dl}	
\end{equation}
\noindent
where $x=1+z$, $g = \Omega_{\rm m} /
\Omega_{\rm Q}$, and the function containing the dark energy equation of state
is given by
\begin{equation}
Q[w,z] = \exp
\left [3\int_1^{1+z} dx
\,w(x)/x \right ]\,\,. 
\label{eq:q(x)}
\end{equation}
We have set $c$ and $H_0$ equal to unity in these expressions. We can
approximately linearize this equation about a fiducial $w^{\rm fid}(z)$
through
\begin{eqnarray}
\label{eq:approximation}
D_L[w^{\rm fid}+ \delta w, z] &\approxeq& D_L[w^{\rm fid}, z] + \delta D_L \\
\delta D_L &=& \int_0^z K_w(z,z') \delta w(z') dz' \,\,,
\nonumber
\end{eqnarray}
where the kernel $K_w(z,z')$ is given by the functional derivative of the
luminosity distance with respect to $w(z)$, evaluated about $w^{\rm fid}(z)$
\begin{equation}
K_w(z,z') = \frac{\delta D_L[w^{\rm fid}(z''),z]}{\delta w(z')} \,\,.
\end{equation}
Evaluating the functional derivative for $D_L$ given by Eq. \ref{eq:dl}
gives 
\begin{equation}
K_w(x,x')=
\left\{ 
\begin{array}{l}
\displaystyle -\frac{3x(1+g)^{1/2}}{2x'} 
\int_{x'}^{x} \frac{dy}{y^{3/2}} \frac{Q[w^{\rm fid}, y]}{\left (g+Q[w^{\rm fid}, y] \right )^{3/2}} \\
\displaystyle 0 \quad \qquad \mbox{for} \,\, x < x' 
\end{array}
\right. \,\,
\end{equation}

The accuracy of the linear response approximation was shown in SPB to
be better than that achieved by the SuperNova Acceleration Probe
(SNAP) survey, which is expected to observe about $2000$ Type~1a SNe,
up to a redshift $z \sim1.7$, each year (Aldering
et al.~2002). By binning the supernovae in a
redshift interval of $\sim 0.02$, SNAP is expected to give a relative error
in the luminosity distance of about $\sim 1\%$. In SPB the accuracy of
the linear response approximation was shown to be better than $1\%$,
implying that it can be used conveniently for a SNAP class data.
However, there are indeed finite departures from the linear
approximation and one must be careful in applying it to real data.
For the purposes of this paper we use this approximation for
convenience, since it enables us to regularize the free form
estimation of the equation of state analytically. We shall apply the
linear response approximation only to those models for which the
departures from the exact expression are small. The usefulness of the
linear approximation, however, can be improved by employing an
iterative method described below.

\begin{figure}
\vbox{\center{\centerline{
\mbox{\epsfig{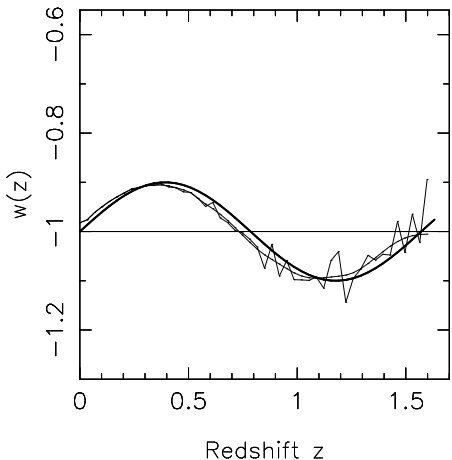}}
}}}
\caption{A free form reconstruction of $w(z)=-1 + 0.2\, \sin\,4z$ without 
added noise. For this reconstruction we have made the assumption that
the numerical noise grows linearly with the distance. The thick solid
line shows the input model, the wiggly line is the reconstruction
without any regularization and the solid line shows reconstruction
with regularization. Small departures form the true equation of state
are due to numerical noise and inexactness of the linear response
approximation as before.}
\label{fig:figure2}
\end{figure}
\section{Free form reconstruction of $w(z)$}
\label{sec:direct}
For the purposes of this paper we simulate the luminosity distance
$D_L$ at a large number, $N_{\rm dat}$, of uniformly distributed
redshifts ${z_i}$, up to a maximum redshift $z=1.7$. We assume a
Gaussian noise equivalent to $1\%$ relative error in the distances
with $\sigma_i$ as the variance.  As a first approximation we fit the
simulated data to a constant $w$ model to obtain $w=w_0$. If
$\Omega_m$ is known to a good accuracy this will give us a first good
approximation to the equation of state. We then linearize the
luminosity distance around $w^{\rm fid} = w_0$.  The difference
between the given noisy estimate of luminosity distance and the best
fit $D_L$ as obtained from $w=w_0$ model gives us the residuals
$\delta D_L$. 

For modelling purposes we consider a discrete version of
Eq. \ref{eq:approximation}
\begin{equation}
\delta D_L(z_i) \approxeq \delta z \sum_{j=1}^{N_{\rm bin}} K_w(z_i,z'_j) 
\delta w(z'_j)\,\,. 
\end{equation}
\noindent
To quantify departures from the constant $w=w_0$ we consider the
equation of state to be given in $N_{\rm bin}$ uniformly distributed
redshift bins ${z'_k}$ with values $w_k = \delta w(z_k')$, with a
redshift spacing $\delta z$. The number and placement of bin positions
for $\delta w$ is such that the the equations generated by the maximum
likelihood procedure, to be described below, yield a unique
solution. The value of $\delta w$ at the position of the farthest
given distance cannot be inferred from the data, since it has no
effect on any of the distances. In practice we also exclude those bins
that are close to the farthest redshift for reconstruction, since the
reconstruction is too noisy in those bins. This particular expansion
is especially convenient since if we have prior knowledge about the
range of $w(z)$ then it is extremely easy to code that into the
reconstruction procedure.

We define the normalized vector $\mathbf {d}
\equiv \{\delta D_L(z_i)/\sigma_i \}$. Similarly we define $ \mathbf
{w} \equiv
\{\delta w(z'_i)\}$ and $\mathcal{K} \equiv \{ \delta z K(z_i,z'_j)/
\sigma_i \}$, where
$\mathcal{K}$ is a $N \times M$ matrix. 
In terms of these quantities a maximum likelihood reconstruction is equivalent
to the standard procedure of minimizing the $\chi_s^2$ function 
\begin{equation}
\chi_s^2 = ({\mathbf d} - {\mathcal
K}{\mathbf w})^{T}({\mathbf d} - {\mathcal K}{\mathbf w})
\label{eq:chisquare}
\end{equation}
with respect to $\bf w$. This can be done analytically to give
\begin{eqnarray}
\mathbf w = ({\mathcal K}^{\rm T}{\mathcal K})^{-1}{\mathcal K}^{\rm T} 
{\mathbf d}\,\,,
\label{eq:linsolution}
\end{eqnarray}
as the required estimator for $\bf w$. In general
Eq~\ref{eq:linsolution} gives a very noisy estimate for the equation
of state. This is due to the fact that too
many parameters are being estimated so the estimator fits most
of the noise as well, and this lack of resolution is a general
feature of the free form fitting (Sivia, 1996).  
This formulation gives the Fisher matrix 
trivially as 
\begin{equation}
F = K^TK \,\,.
\label{eq:fisher} 
\end{equation}
Huterer \& Starkman, (2003) diagonalize the Fisher matrix and find
that only a few eigenvectors are well determined. They expand the
equation of state in terms of the eigen vectors of the Fisher matrix
to obtain an approximate form for the equation of state by truncating
the series after the first few well determined eigenvectors.  They
note that truncating the series biases the estimation.  They find that
the badly determined eigen vectors are precisely those that peak at
high redshift, therefore, throwing away those eigenvectors biases the
equation to state to zero at high redshift. In the next section we 
describe another approach that does not have this problem and is better
able to represent the equation of state at all redshifts.

\begin{figure}
\vbox{\center{\centerline{
\mbox{\epsfig{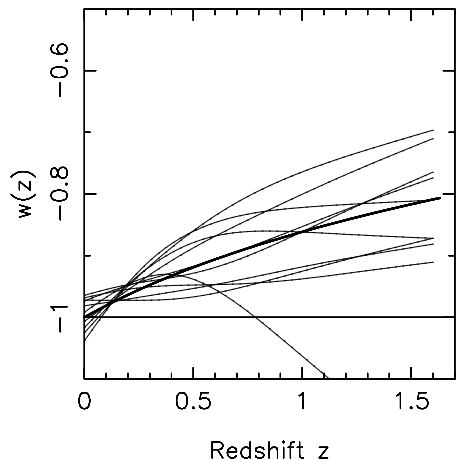}}
}}}
\caption{Regularized reconstruction of the same model with $1\%$ Gaussian 
noise added to the distances. Different curves show reconstructed 
equation of state for different realization of noise. As expected, the
reconstruction is better at the low redshift and has a large scatter
at high redshifts. 
}
\label{fig:figure3}
\end{figure}

\subsection{Regularization}

The expansion described above does not guarantee smoothness or
continuity.  In fact, for large $N_{\rm bin}$ the solution fits most
of the noise as well, and therefore gives a very small value of
$\chi^2$. Since an acceptable model would give $\chi^2 \sim N_{\rm
dat}$ we might wish to modify the best fit $\mathbf w$ to ensure some
smoothness. As noted above, the amount of information contained even
in a SNAP class experiment is too little to adequately constrain
$w(z)$.  Additional information in terms of continuity and smoothness
constraints tend to fuzz information across the bins and reduces the
degrees of freedom of the free form fitting function. In regions where
the data is not discriminatory enough, the derived equation of state
tends to extrapolate from the well constrained regions and can provide
useful information.  As an obvious warning we note that in general
this is a dangerous thing to do since it might bias the estimator, or
even create spurious signal. In our view these disadvantages are
outweighed by the fact that such a scheme would gives us, at least, a
fighting chance to infer the behaviour of $w(z)$ in a way that is
independent of the specific form of $w(z)$ chosen to fit the data.
Regularization also reminds us that we are \emph{explicitly} assuming
smoothness for $w(z)$, rather than sneaking it in the form of smooth
fitting functions.

To ensure continuity and smoothness we modify the $\chi^2$ above to
\begin{eqnarray}
\chi^2 &=& \chi_s^2 +
\lambda \sum_{i=1}^{N-1}(w_{i+1} - w_i)^2\\ \nonumber
&& +\beta \sum_{i=1}^{N-2}(w_{i+2} -2 w_{i+1} + w_i)^2
\end{eqnarray}
The added terms ensure that the first and the second derivative of $w$
remain small.  The weights $\lambda$ and $\beta$ present the problem
that in the absence of prior information about the first and the
second derivatives of $w$ it is difficult to decide how to set them in
such a way that the solution does not get biased significantly.  If
$\lambda$ is set too large then the preferred solution is a constant
$w$, and if $\beta$ is set too large then the preferred solution is a
linear function of $z$. If the data is of good quality, and a linear
or a constant solution are not good approximations to the true $w(z)$,
then the inclusion of these terms will drive the solution away from
the ideal solution and will drive the usual $\chi_s^2$ to unrealistic
values. Therefore, our choice for $\lambda$ and $\beta$ ensures that
the value of $\chi_s^2$ should lie in the range $N_{\rm dat} -
\sqrt{2N_{\rm dat}} <\chi_s^2 < N_{\rm dat} +
\sqrt{2N_{\rm dat}}$. 

It is convenient to write the above equation in the form
\begin{eqnarray}
\chi^2 &=& \chi_s^2 +
{\lambda \,\mathbf w (\mathcal F^T \mathcal F) \mathbf w} 
+ {\beta \, \mathbf w (\mathcal S^T  \mathcal S) \mathbf w}
\end{eqnarray}
where $\mathcal F$ is a $(N_{\rm bin}-1) \times N_{\rm bin}$ matrix
and is non-zero for
\begin{equation}
\mathcal F_{ij} =
\left\{ 
\begin{array}{rl}
-1& \quad \qquad \mbox{for} \,\, i=j \\
 1& \quad \qquad \mbox{for} \,\, i=j-1 
\end{array}
\right. \,\,.
\end{equation}
\noindent
and $\mathcal S$ is a $(N_{\rm bin}-2) \times N_{\rm bin}$ matrix and is non-zero 
for
\begin{equation}
\mathcal S_{ij} =
\left\{ 
\begin{array}{rl}
-1& \quad \qquad \mbox{for} \,\, i=j \\
 2& \quad \qquad \mbox{for} \,\, i=j-1 \\
-1& \quad \qquad \mbox{for} \,\, i=j-2 
\end{array}
\right. \,\,.
\end{equation}
In terms of this new regularized $\chi^2$ the solution Eq~\ref{eq:linsolution}
is modified to 
\begin{eqnarray}
\mathbf w = ({\mathcal K}^{T}{\mathcal K} +\lambda {\mathcal F}^T{\mathcal F} + \beta {\mathcal S}^{T}{\mathcal S} 
)^{-1}{\mathcal K}^{\rm T} {\mathbf d}\,\,,
\label{eq:newlinsolution}
\end{eqnarray}
For further details about regularization and some interesting remarks
about the connection between the previous expression and an optimal
Weiner filter the reader is referred to Chap~18 of Press et al. (1992).
\begin{figure}
\vbox{\center{\centerline{
\mbox{\epsfig{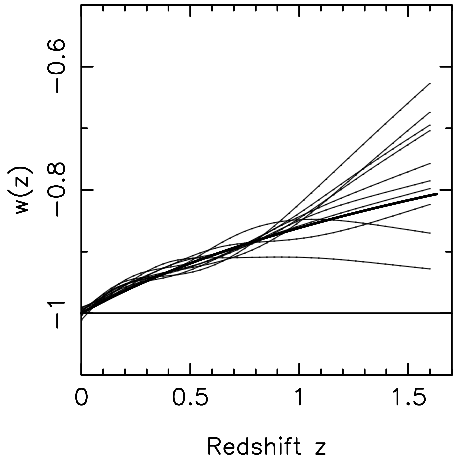}}
}}}
\caption{Reconstruction for the same model and same realizations as in Fig~\ref{fig:figure3} but with $10$ times less noise. Reconstruction works very well
at the low redshift. At redshifts $z > 1$ the reconstruction is slightly 
biased, consistent with Fig~\ref{fig:figure1}, due to inexactness of the 
linear response approximation. 
}
\label{fig:figure4}
\end{figure}

\section{Numerical Investigations}
For our numerical explorations we adopt the following model for 
the dark energy as the input equation of state
\begin{equation}
w(z) = w_0 + w_1 \ln(1+z)\,\,,
\end{equation}
\noindent
since it has the nice property that $w_0$ and $w_1$ are the present
day value of $w(z)$ and its first derivative respectively.  Since this
model has finite departures from the equivalent linear model it also
serves to illustrate how well a regularized estimation gives
information about departures from a simpler linear model.  The
function $Q[w,z]$ defined in Eq~\ref{eq:q(x)} is calculated
analytically to obtain
\begin{equation}
Q[w,z] = x^{3w_0}\exp [\frac{3w_1}{2} \ln^2(1+z) ]
\end{equation}
Since the linear response approximation is an expansion about a given
$\Omega_m$, we shall assume its value to be exactly known.  We
generate the simulated data in $N_{\rm dat} = 50$ bins up to a maximum
redshift $z=1.7$. For reconstruction purposes we employ $N_{\rm bin} =
48$ bins for placing the binned $w(z)$. As remarked earlier, we do not
reconstruct $w(z)$ above a redshift of $z \sim 1.6$ due to poor
resolution.

The unregularized estimator given in Eq~\ref{eq:linsolution} picks up
numerical noise even when the distances are exactly given. To show how
unstable such an estimation is we contrast it with the regularized
estimation in Fig~\ref{fig:figure1}, which shows the reconstructed
equation of state for $w_0=-1$ and $w_1=0.2$ model without added noise
(but with small numerical noise due to inexact mathematical
operations). The un-regularized reconstruction follows the true model
quite closely but picks up numerical noise and appears noisy.  The
regularized reconstruction, on the other hand, follows the true
equation of state quite closely. The tiny departures from it are due
to the fact that the linear response approximation is not exact. This
clearly shows that a reconstruction without regularization is
ineffective even when the statistical noise is absent. 

Since the unregularized estimator is extremely sensitive to noise, we
can improve its performance by taking into account the inevitable
numerical noise due to inexact mathematical operations to make the
comparison fair. To a good approximation the numerical noise
grows linearly with the distance. If we normalize $\mathbf {d} =
\{\delta D_L(z_i)/\sigma_i
\}$ and $\mathcal{K} = \{
\delta z K(z_i,z'_j)/ \sigma_i \}$ with $\sigma_i
\propto D_l(z_i)$ then we find that the unregularized estimation is
stabilized to some extent. To illustrate this effect, and to show how
the reconstruction works for a model that shows strong departures from
a straight line we also reconstruct the equation of state $w(z)= -1.0
+ 0.2\,\sin\,4z$ in Fig~\ref{fig:figure2} without added statistical
noise.  We find that the reconstruction works better than before for
the unregularized case, but the regularized estimation works better in
both the cases.

To quantify how the regularized estimator works for \emph{noisy} data we add
a $1\%$ Gaussian noise to the previous model ($\sigma_i = .01 \times
D_L(z_i)$). We fixed the weights $\lambda$ and $\beta$ to ensure a
reasonable $\chi_s^2$ for this experiment. Fig~\ref{fig:figure3} shows
the outcome for $10$ realizations of the noisy data. The reconstructed
equation of state typically has a scatter of about $\sim 0.2$ at low
redshift. At large redshifts the reconstruction is poor but remain
unbiased.  To see the effect of decreasing noise we lower it by ten
times for Fig~\ref{fig:figure4}. The reconstruction now works much
better, as expected. The scatter at the large redshift is still
substantial, however, the equation of state is reconstructed very well
up to $z \sim 0.7$. We also find that the reconstruction recovers the
shape of $w(z)$, even up to large redshifts in many cases. In part
this success is due to the fact that regularization extrapolates from
the well determined low redshift $w(z)$.

\subsection{Noise Estimation}
Quantifying errors on the reconstructed equation of state is not
simple in this approach. One way to do this is to use the
reconstructed equation of state to calculate the Fisher matrix in
Eq~\ref{eq:fisher}.  The inverse of that would yield the required
covariance matrix. This would be equivalent to assuming a Gaussian
distribution for the estimator in Eq~\ref{eq:newlinsolution}. Our
numerical experiments show that the distribution of equation of states
generated for different realizations of noise by the regularized
estimator is not close to Gaussian. In fact, as Fig~\ref{fig:figure3}
shows, the curves fill the space in a very complex fashion so the
distribution of $w$ values at any given redshift are not well
approximated by the Gaussian distribution.  To take into account this
effect we generate $1000$ realizations of data and plot the envelope
of all the curves obtained from different realizations. The scatter
is, of course, more concentrated around the true equation of state.
The result is shown in Fig~\ref{fig:figure5}, where for comparison we have
plotted the same for the case where $\beta$ is kept equal to zero. The
figure shows that the when $\beta \ne 0$ the noise goes down in the
estimation at all redshifts. This is what we would expect since this
term adds information about the second derivative in the estimation.

The envelope for $\beta = 0$  shows a weak sweet spot at $z
\sim 0.1$ while the one with a non-zero $\beta$ the sweet spot
becomes stronger and moves to $z \sim 0.2$. In fact, if we consider
the Fisher matrix of the unregularized estimation in
Eq~\ref{eq:fisher} we find that no sweet spot appears anywhere. This
is indeed as it should be since the $w$ values at small redshifts
affect the distances at all the higher redshifts, so should be better
constrained. If we increase $\beta$ the sweet
spot moves to $z \sim 0.3$ and becomes stronger. This behaviour
suggests that the sweet spot is, in general, an artifact of the
particular method employed in the estimation. The demand that the
$w(z)$ have a small second derivative cannot fit the data at all
redshifts adequately, so it develops an anchor at an intermediate
redshift, as is seen in Fig~\ref{fig:figure3}.  Huterer \&
Starkman~(2003) express similar views on the origin of the sweet spot
in their paper.

\begin{figure}
\vbox{\center{\centerline{
\mbox{\epsfig{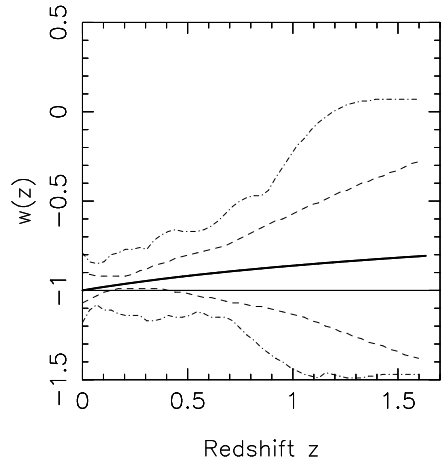}}
}}}
\caption{Here we quantify the scatter in the regularized estimator
by plotting the envelope enclosing $1000$ realizations of noisy
data. The dot-dashed curve has $\lambda$ set to a fixed value
to give $\chi^2 \sim N_{\rm dat}$ and $\beta = 0$. The dashed curve
has both $\lambda$ and $\beta$ set to fixed values. Imposing
two regularizing conditions decreases the scatter and improves
the performance at low redshift. 
}
\label{fig:figure5}
\end{figure}

\section{Conclusions}

We have shown that the limitations of the free form estimation of dark
energy can be overcome by simple regularization, which substantially
improves its performance. Our regularizing terms incorporate
smoothness of the equation of state and its first derivative, but can
be easily generalized to take into account further information about
the equation of state. We have explicitly constructed an analytic form
for the regularized estimator for the equation of state by employing
the recently introduced linear response approximation which allows the
maximum likelihood estimation to be performed analytically. Although the
linear response approximation is not perfect, its applicability can
be extended by iteration. The major uncertainty in the determination
of the properties of dark energy is the present day density in the
form of pressureless dark matter. The results in this paper assume
this density to be known but can be easily generalized to take this
into account properly. Due to its analytical simplicity and due to the
fact that by construction it is unbiased, the regularized free from
estimation is superior to all others. However, we do find that
artifacts of regularization appear in the form of sweet spots. It is
fair to say that at the level of accuracy that a SNAP class experiment
will achieve it seems unlikely that this method will give any more
information than that given by simple polynomial fits to the equation
of state. Since this method is well suited for combining supernovae
data with other cosmological tests it might yet prove to be a more useful
way of constraining the properties of dark energy.

\section{Acknowledgments}

TDS acknowledges financial support from PPARC. We thank Sarah Bridle,
Ofer Lahav and Shiv Sethi for useful discussions.

\newcommand{\ea}{et~al.\ }
\def\bb#1#2#3#4#5#6#7{\bibitem[\protect\citename{#2 }#3]{#1}#4, #3,
#5, #6, #7}
\def\bbprep#1#2#3#4#5{\bibitem[\protect\citename{#2 }#3]{#1}#4, #3, #5}
\def\prl{Phys.\ Rev.\ Lett.}
\def\pr{Phys.\ Rev.}
\def\pl{Phys.\ Lett.}
\def\np{Nucl.\ Phys.}
\def\prp{Phys.\ Rep.}
\def\rmp{Rev.\ Mod.\ Phys.}
\def\cmp{Comm.\ Math.\ Phys.}
\def\mpl{Mod.\ Phys.\ Lett.}
\def\apj{Ap.\ J.}
\def\apjl{Ap.\ J.\ Lett.}
\def\aap{Astron.\ Ap.}
\def\cqg{Class.\ Quant.\ Grav.} 
\def\grg{Gen.\ Rel.\ Grav.}
\def\mn{MNRAS}
\def\ptp{Prog.\ Theor.\ Phys.}
\def\jetp{Sov.\ Phys.\ JETP}
\def\jetpl{JETP Lett.}
\def\jmp{J.\ Math.\ Phys.}
\def\zpc{Z.\ Phys.\ C}
\def\cupress{Cambridge University Press}
\def\pup{Princeton University Press}
\def\wss{World Scientific, Singapore}
\def\oup{Oxford University Press}
\def\asj{Astron.~J}
\def\imp{Int.\ J.\ Mod.\ Phys.}

\end{document}